\documentclass[12pt,british,english]{iopart}

\usepackage[T1]{fontenc}
\usepackage[latin9]{inputenc}
\usepackage{graphicx}

\makeatletter
\usepackage{iopams}
\usepackage{setstack}

\usepackage{cite}

\@ifundefined{showcaptionsetup}{}{%
 \PassOptionsToPackage{caption=false}{subfig}}
\usepackage{subfig}
\makeatother

\usepackage{babel}
\begin{document}
\selectlanguage{british}%

\title{Simulations of photon detection in SiPM number-resolving detectors}

\author{L Dovrat, M Bakstein, D Istrati and H S \foreignlanguage{english}{Eisenberg}}

\address{Racah Institute of Physics, the Hebrew University of Jerusalem, Israel}

\ead{liat.dovrat@mail.huji.ac.il}
\selectlanguage{english}%
\begin{abstract}
Number-resolving single photon detectors are essential for the implementation
of numerous innovative quantum information schemes. While several
number-discriminating techniques have been previously presented, \foreignlanguage{british}{the
Silicon Photo-Multiplier (SiPM) detector is a promising candidate
due its rather simple integration in optical setups. On the other
hand, the photon} statistics obtained with the SiPM detector suffer
from inaccuracies due to inherent distortions which depend on the
geometrical properties of the SiPM\foreignlanguage{british}{. We have
simulated the detection process in a SiPM detector and studied these
distortions. We use results from the simulation in order to interpret
experimental data and study the limits in which available models prevail. }
\end{abstract}
\selectlanguage{british}%


\maketitle

Photon-number discriminating detectors have been the subject of great
interest over the past few years. Number resolving capabilities are
the key to several quantum state preparation schemes~\cite{KokDowling2002,LeeDowling2002,GaoDowling2010}
as well as in the implementation and analysis of quantum computation
schemes~\cite{Pittman2002,Gisin2004,KLM2001,Okamoto2009,Franson2002}.
Photon-number resolution also enables the direct measurement of a
state's photon-number statistics, from which non-classical properties
as well as classical-to-quantum transitions can be studied and characterized~\cite{Waks2004,Waks2006,Short1983}.

Standard single photon detection techniques cannot resolve the number
of photons, and number resolution is generally obtained by combining
several number-insensitive detectors in a cascade or an array-like
structure. One of the first photon-number resolving detectors introduced
was the Visible Light Photon Counter (VLPC)\cite{VLPC_Takeuchi1999}.
The VLPC has a high detection efficiency, but requires cryogenic
cooling. Superconducting devices which provide number resolution are
available but also require low working temperatures\cite{TES_Cabrera1998,Fujiwara2006,Hadfield2005,Rosenberg2005,DaulerBerggren2009}.
Room-temperature based solutions using standard single \foreignlanguage{english}{avalanche
photodiode (APD}) detectors offer limited number-resolution\cite{Kardynal2008},
while other solutions based on spatial and temporal multiplexing are
experimentally demanding~\cite{Kok2001,Paul1996,Achilles2004,Fitch2003}. 

A promising approach is provided by the Silicon Photo-Multiplier (SiPM)\cite{Bondarenko1998}.
The SiPM is composed of multiple silicon avalanche photo-diodes arranged
on a single substrate. Each detection element acts as an independent
avalanche photo-diode in Geiger-mode, which can absorb a photon and
generate a confined electric discharge. Signals from all the detecting
elements are combined to a single readout port, so that the intensity
of the output signal is proportional to the number of impinging photons.
The SiPM detector offers a very good number resolution, operates at
room temperature and is easily integrated in optical setups.

\selectlanguage{english}%

When using SiPMs, three main factors affect the measured photon statistics.
These effects are the relatively low detection probability, determined
by the quantum efficiency and geometrical configuration, internal
noise caused by thermal excitations resulting in false detection,
and optical crosstalk (CT) in which a photon created by carrier relaxation
in one detection element is detected in a neighboring element~\cite{Buzhan2006,Dolgoshein2006}.
There exist several models~\cite{Afek2009,Akiba2009,Eraerds2007}
which take these phenomena into account. These models aid in the interpretation
of experimental results and allow the reconstruction of the original
photon-number statistics. However, these available models handle the
CT effect in a rather limiting manner, either by limiting the CT probability,
limiting the overall number of CT events, or limiting the number of
CT stages (ignoring crosstalk events generated by crosstalks). Since
these models were inaccurate in describing our experimental results,
\foreignlanguage{british}{we have written a computational model which
simulates the detection process in SiPM detectors. Using this model,
we determine the conditions under which available models are accurate,
and examine the limits imposed by their assumptions with regards to
interpretation of experimental data.}

\selectlanguage{british}%

In the model, the detector is represented by a two-dimensional lattice
where each cell represents an element of the detector. The impinging
photons are distributed uniformly across the lattice. When a photon
approaches a certain cell, the cell is triggered with some probability,
$\eta$, the detection efficiency. If triggered, each of the cell's
nearest neighbours can also be triggered with a probability $\epsilon_{nn}$,
the optical crosstalk probability. These new triggered elements can
continue to trigger their remaining nearest neighbours. This process
continues until no new cells are triggered. The dead time of the detection
elements in available SiPM detectors is longer than the photon propagation
time between cells~\cite{Buzhan2006}, so our model allows each element
to only be triggered once in the process. The statistics were obtained
on a $10\times10$ lattice with 100 elements, which corresponds to
a typical commercially available SiPM detector (\emph{Hamamatsu Photonics},
\foreignlanguage{english}{S10362-11-100U}~\cite{Hamamatsu_url}).
Nevertheless, the conclusions can be inferred to other sized sampled
by translating the absolute number of detections to the fraction of
occupancy (20 photons in a 100 pixel detector are equivalent to 80
photons in a 400 pixel detector). The samples are not entirely scalable
in size since the number of cells along the borders does not scale
with the number of elements. However, simulations performed on different
sized samples showed that this has little effect on the overall results
in the range of the experimental parameters.

\begin{figure}
\centering{}\subfloat[\label{fig:1n_0.05eps}]{\begin{centering}
\includegraphics[width=0.2\columnwidth]{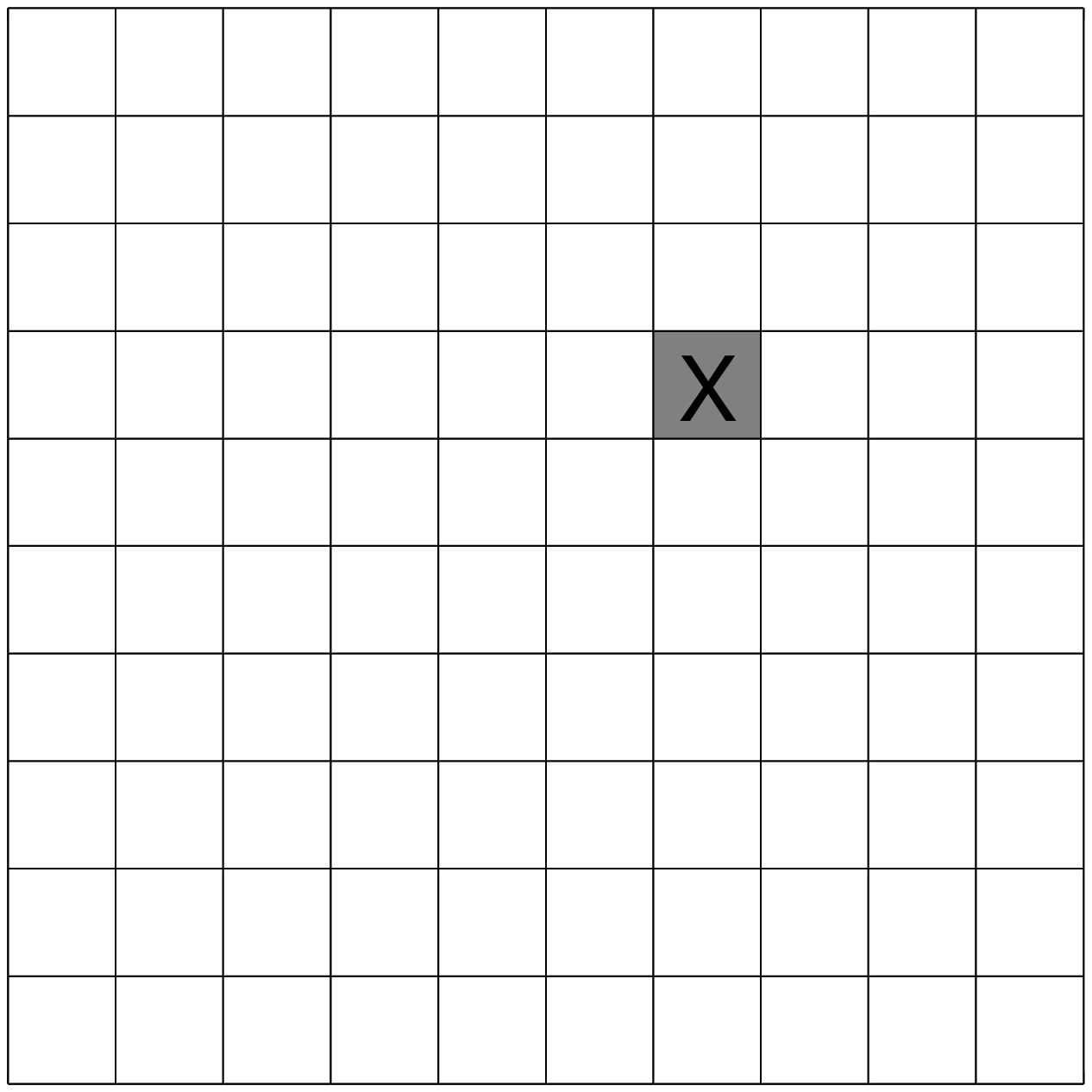}
\par\end{centering}

}\subfloat[\selectlanguage{english}%
\label{fig:1n_0.5eps}\selectlanguage{british}
]{\begin{centering}
\includegraphics[width=0.2\columnwidth]{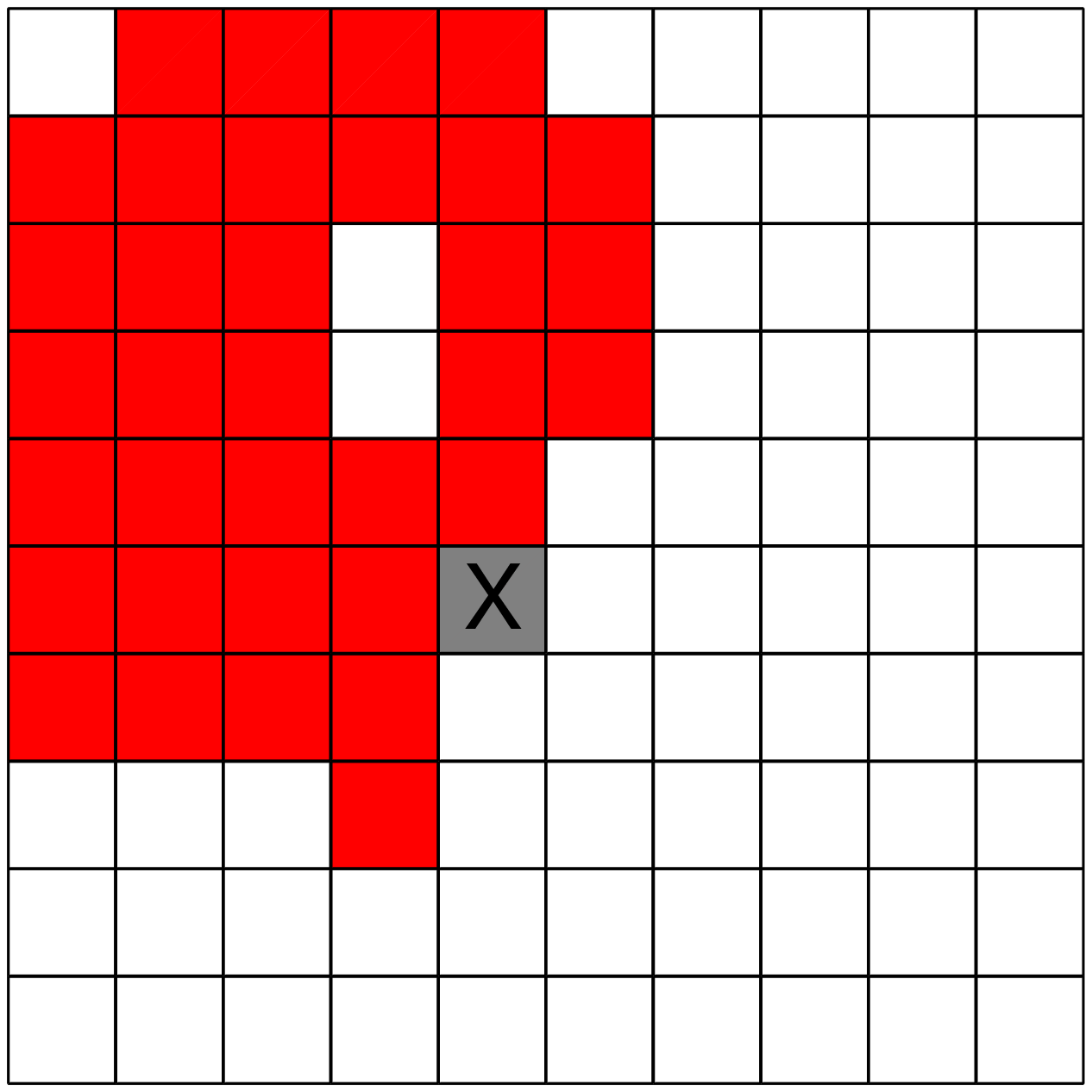}
\par\end{centering}

}\\
\subfloat[\label{fig:60n_0.05eps}]{\begin{centering}
\includegraphics[width=0.2\columnwidth]{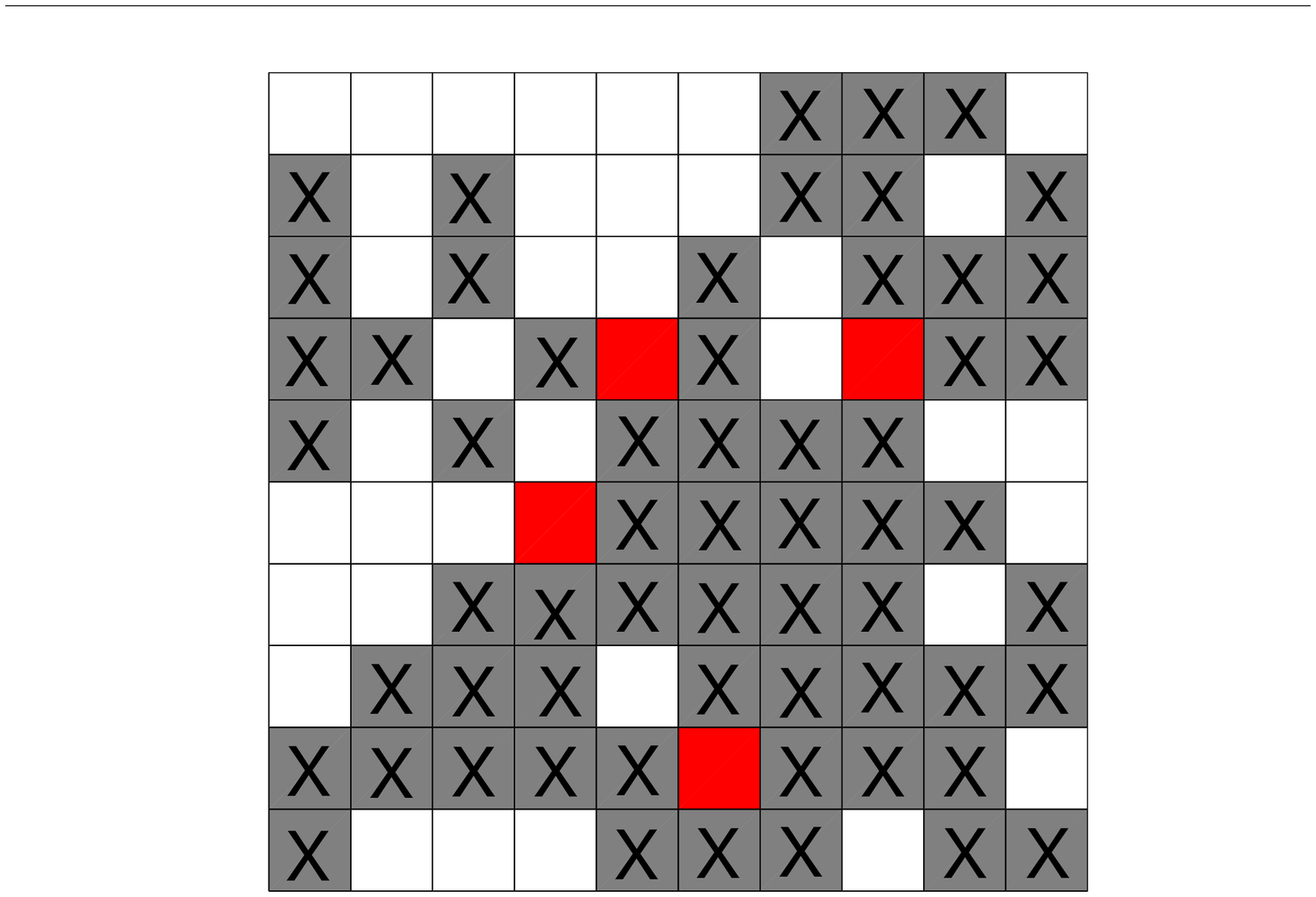}
\par\end{centering}

}\subfloat[\label{fig:60n_0.5eps}]{\begin{centering}
\includegraphics[width=0.2\columnwidth]{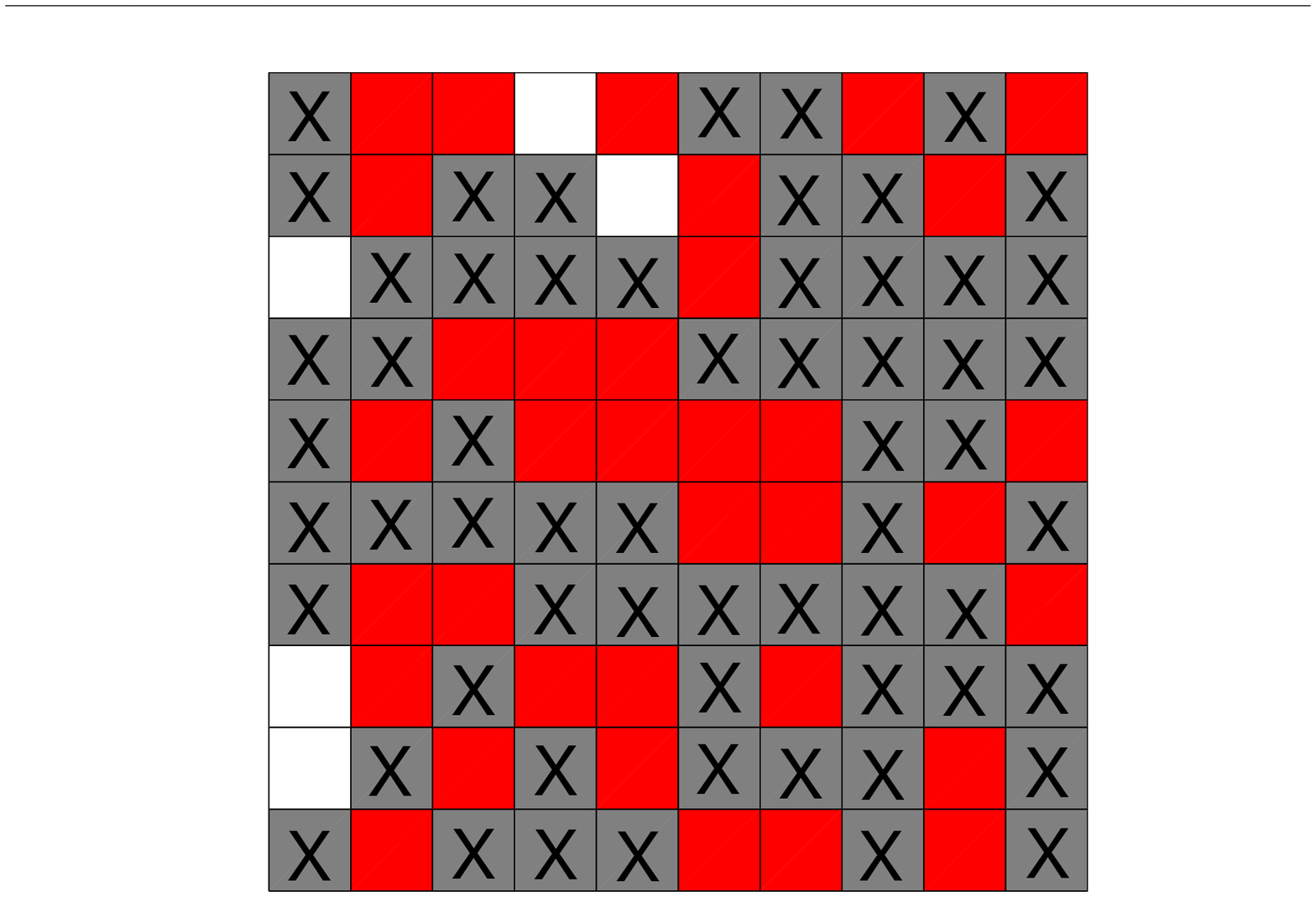}
\par\end{centering}

}\caption{\selectlanguage{english}%
Representative simulation runs of the crosstalk evolution. \foreignlanguage{british}{The
initially triggered pixels are marked with a cross and coloured grey,
and the crosstalk triggered pixels are shown in red.} (a) $N_{trg}=1,\,\epsilon_{nn}=0.05$,
0 crosstalks (b) $N_{trg}=1,\,\epsilon_{nn}=0.5$, 34 crosstalks (c)
$N_{trg}=60,\,\epsilon_{nn}=0.05$, 4 crosstalks\foreignlanguage{british}{
(d) }$N_{trg}=60,\,\epsilon_{nn}=0.5$\foreignlanguage{british}{,
35 crosstalks.}\label{fig:Simulation-examples of one pixel} \selectlanguage{british}
}
\end{figure}

\selectlanguage{english}%
We begin our discussion with the optical crosstalk process. \foreignlanguage{british}{Figure~\ref{fig:Simulation-examples of one pixel}
shows representative runs which portray the crosstalk evolution. The
detector is initially triggered by $N_{trg}$ detections of the impinging
photons. These triggers initiate a crosstalk process which results
in the triggering of additional cells. The number of crosstalk events
shown in these examples was chosen so that it corresponds to the average
number of CTs produced over multiple runs. As expected, the number
of triggered cells increases with the crosstalk probability, $\epsilon_{nn}$.
In figures~\ref{fig:1n_0.5eps} and~\ref{fig:60n_0.5eps}, we present
an example of the effect the finite number of elements has on the
number of generated CTs. Increasing the value of $N_{trg}$, beyond
some critical limit, does not result in additional crosstalk triggers.
The majority of cells attempt to trigger neighbouring cells which
have already been triggered, and do not contribute to the evolution
of the process.}

\selectlanguage{british}%

\begin{figure}
\begin{centering}
\includegraphics[width=86mm]{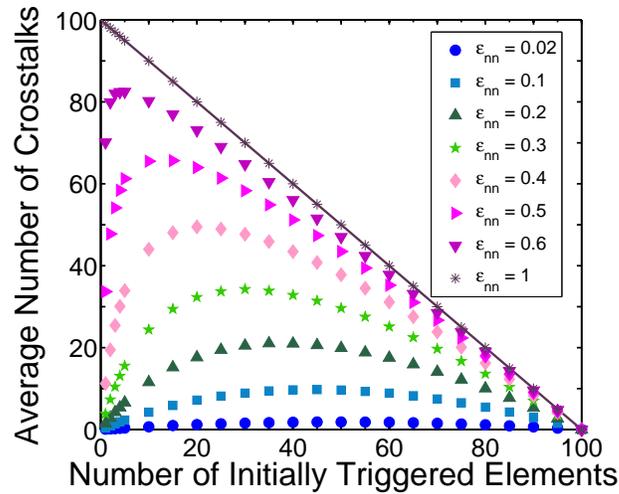}
\par\end{centering}

\caption{\selectlanguage{english}%
The average number of crosstalks events as a function of the initial
number of triggers for different values of $\epsilon_{nn}$. The number
of crosstalk events increases monotonically with the number of initial
triggers until some peak value is reached, after which the number
begins to decrease due to finite size effects. For $\epsilon_{nn}=1$,
the detector is saturated regardless of the number of original triggers.
The solid line is the function $N_{CTs}=N_{elements}-N_{trg}$.\label{fig:NumCTs vs Nhits}\selectlanguage{british}
}
\end{figure}

\selectlanguage{english}%
The limited number of cells can cause the number of crosstalk events
to decrease rather than increase with the addition of more initial
triggers. This result is shown in \foreignlanguage{british}{figure~\ref{fig:NumCTs vs Nhits}.
When the average number of triggered elements is small, the number
of CTs grows linearly with $N_{trg}$. As $N_{trg}$ is increased,
the effective number of neighbouring cells, which can also be triggered,
is reduced as many of these neighbouring cells have already been triggered.
As a result, the average number of crosstalk events deviates from
linearity and starts decreasing after reaching some peak value. Notice
that this effect, which is caused by the finite size of the detector,
comes into play even at relatively low numbers of detected photons.}

\selectlanguage{british}%

\begin{figure}
\begin{centering}
\includegraphics[width=86mm]{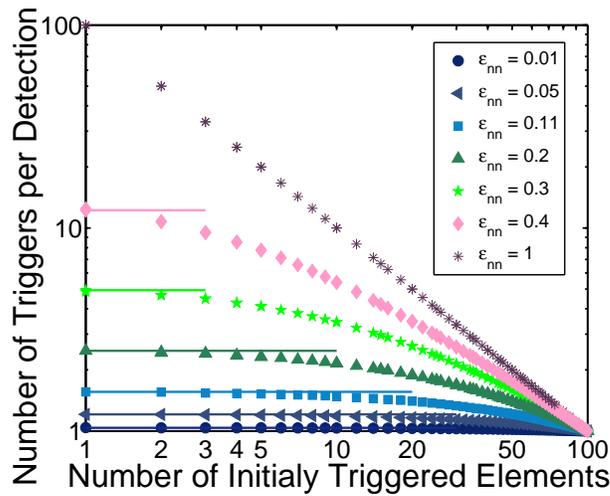}
\par\end{centering}

\caption{\selectlanguage{english}%
The average number of triggered elements generated by a single detection.
This value is constant and begins to decrease when the finite number
of elements begins to impose a limitation. \label{fig:Cluster size}\selectlanguage{british}
}
\end{figure}

In order to identify the critical point where finite size effects
begin to dominate the crosstalk process, we consider the average number
of triggered elements generated by one initial trigger.\foreignlanguage{english}{
In a detector of infinite size the crosstalk expansion is only suppressed
by the probability for crosstalk. When the number of elements is limited,
the progression is also suppressed due to overlaps with previously
triggered cells. The average number of triggered elements should therefore
be constant in the absence of finite size effects and should decrease
when the finite size begins to impose a limitation. }Figure~\ref{fig:Cluster size}
shows that for large crosstalk probabilities, the finite size imposes
an immediate restriction, whereas the process expands rather freely
when the crosstalk probability is below $\epsilon_{nn}\approx0.025$.

\begin{figure}
\begin{centering}
\includegraphics[width=86mm]{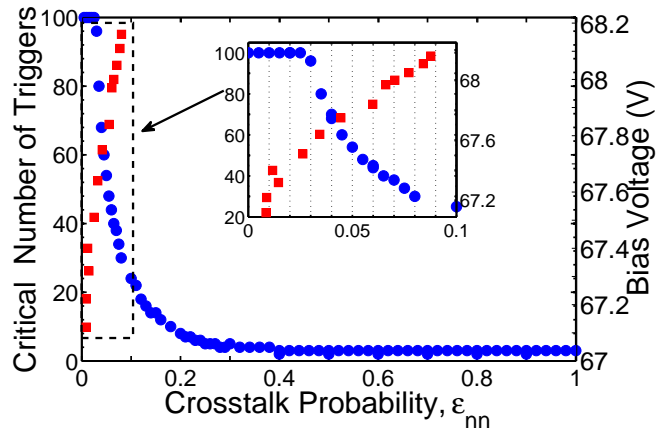}
\par\end{centering}

\caption{\selectlanguage{english}%
The critical number of detected photons above which finite size effects
become significant as a function of the crosstalk probability (left
axis, blue circles), and typical crosstalk values, $\epsilon_{nn}$,
as we measured over a range of bias voltages (right axis, red squares).
The measurements were conducted using a SiPM detector (\emph{Hamamatsu
Photonics}, S10362-11-100U).\label{fig:CT size effect}\selectlanguage{british}
}
\end{figure}

\selectlanguage{english}%
We define the \foreignlanguage{british}{critical number of initial
triggers above which the size effects must be considered as the point
where the average number of elements triggered by a single detection
decreases by more than 10\% compared to the unaffected value. }In
figure\foreignlanguage{british}{~\ref{fig:CT size effect} we show
this critical value as a function of the crosstalk probability, $\epsilon_{nn}$.
The number of crosstalk-triggered cells increases with the value of
$\epsilon_{nn}$, and the value of the critical value is therefore
lower. Figure~\ref{fig:CT size effect} also presents typical crosstalk
values we experimentally measured for a range of bias voltages. The
higher the bias voltage, the higher the gain and the crosstalk probability~\cite{Buzhan2006}.
For the typical range of experimental values, the maximal number of
detected photons for which finite size effects are not significant
is in the range 20--100. These numbers are significantly higher than
the number of photons observed in previously reported works~}\cite{Afek2009,Akiba2009,Eraerds2007,Kalashnikov2011}\foreignlanguage{british}{.
This explains why the modelling of the crosstalk effect without geometrical
features of the detector was sufficient in these works. }

\selectlanguage{british}%

\begin{figure}
\begin{centering}
\includegraphics[width=86mm]{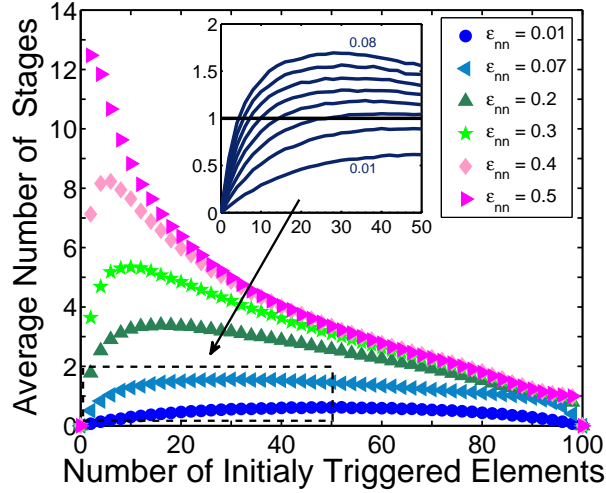}
\par\end{centering}

\caption{\selectlanguage{english}%
The number of crosstalk stages as a function of the number of original
triggers for various crosstalk probability values. The inset shows
the number of stages for values of $\epsilon_{nn}$ in the experimental
range 0.01--0.08 in steps of 0.01. \label{fig:Num CT stages}\selectlanguage{british}
}
\end{figure}

We turn now to evaluate the number of crosstalk stages involved in
the process before it stops. The number of stages is an indication
of the amount of cells triggered due to crosstalk generated by crosstalk
events, and is an important factor in the modelling of the crosstalk
process. The average number of CT stages is shown in figure~\ref{fig:Num CT stages}.
For small crosstalk probabilities only a small number of neighbouring
elements is triggered due to crosstalk and the process ceases naturally
after one stage even for large initial values of $N_{trg}$. For large
values of $\epsilon_{nn}$, the number of crosstalks generated by
other crosstalk events increases and we observe a rise in the number
of stages.\foreignlanguage{english}{ It is interesting to note that
}reference~26 had a CT probability equivalent to $\epsilon_{nn}\approx0.025$
and used a one-stage crosstalk model. From figure~\ref{fig:Num CT stages},
we can see that for this value of $\epsilon_{nn}$, the number of
stages indeed does not exceed 1 even for photon numbers higher than
what was detected in this work. On the other hand, we measured a CT
value of $\epsilon_{nn}\approx0.07$ and detected up to 14 photons.
In this regime the number of stages exceeds 1. In fact, a simplistic
1-stage model would result in an error of over 20\% in the evaluation
of the number crosstalk events. The finite size effects are also apparent
in this graph. Similar to the behaviour observed in figure~\ref{fig:NumCTs vs Nhits},
the number of crosstalk stages begins to decrease slowly after reaching
some peak value.

\begin{figure}
\begin{centering}
\includegraphics[width=86mm]{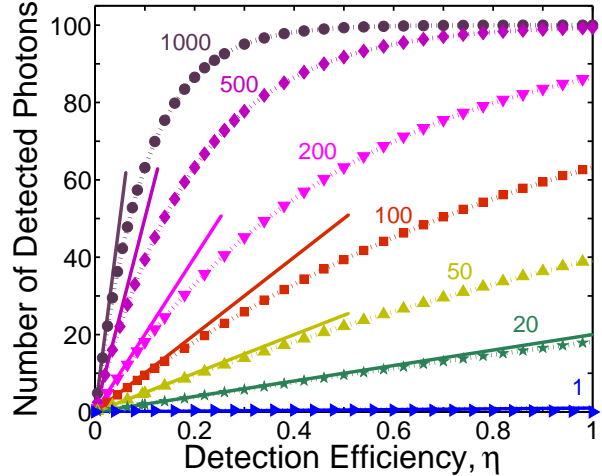}
\par\end{centering}

\caption{\selectlanguage{english}%
The number of detected photons as a function of the detection efficiency
as obtained by our model. \foreignlanguage{british}{Solid lines represent
the linear relation, $\left\langle N_{detected}\right\rangle =\eta\cdot N_{photons}$,
for the respective number of impinging photons}. The value of $N_{photons}$
appears above each plot. The dash lines are function plots of equation~\ref{eq:N_fired}.\label{fig:Dynamic-Range}\selectlanguage{british}
}
\end{figure}

We now consider the second effect which governs the behaviour of SiPM
detectors, the inherent loss mechanism. Apart from the low detection
efficiency, when two or more photons approach the same detection element,
only one avalanche can be generated and the number-resolution is lost.
This non-linear response of the detector, affects the detection probabilities.
The average number of detected photons can be generally written as~\cite{Hamamatsu_url}

\begin{equation}
\left\langle N_{detected}\right\rangle =N_{elements}\times\left[1-\exp\left(-\frac{\eta\cdot N_{photons}}{N_{elements}}\right)\right],\label{eq:N_fired}
\end{equation}
where $N_{photons}$ is the number of impinging photons, $\eta$ is
the detection efficiency and $N_{elements}$ is the number of elements.
In the limit $\eta\cdot N_{photons}\ll N_{elements}$, where finite
size effects can be neglected, the expression reduces to the linear
relation 
\begin{equation}
\left\langle N_{detected}\right\rangle =\eta\cdot N_{photons}.\label{eq:linear Ntrg}
\end{equation}
We use this relation to determine a critical condition below which
the finite number of the detection elements do not affect the detection
probabilities. In figure~\ref{fig:Dynamic-Range} we show the average
number of detected photons as a function of the detection efficiency.
We identify deviations of over 10\% from a linear slope when the fraction
of triggered elements, $\frac{\eta\cdot N_{photons}}{N_{elements}}$,
exceeds 20\%. Interestingly, the deviations from linearity depend
on the detection efficiency and this behaviour may be used in order
to obtain the absolute detection efficiency of the detector.

\begin{figure}
\begin{centering}
\includegraphics[width=86mm]{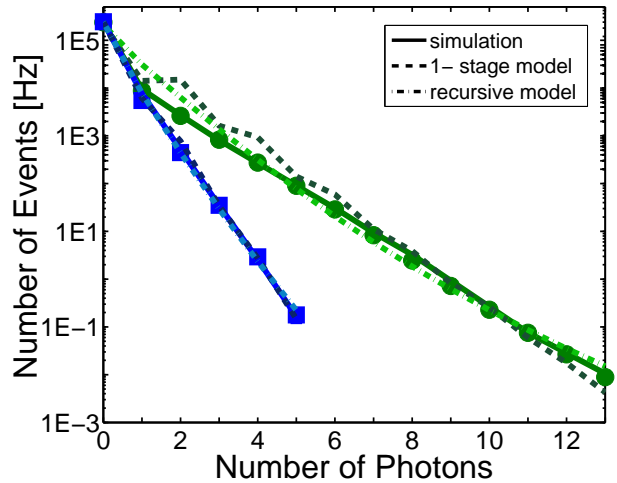}
\par\end{centering}

\caption{\selectlanguage{english}%
Measurements of thermal photon-number statistics taken using a SiPM
detector (\emph{Hamamatsu Photonics}, S10362-11-100U). The experimental
data was fit to the statistics obtained using our computational model
(solid lines), the 1-stage model of reference~26 (dashed line) and
the recursive model of reference~28 (dot-dashed line), with crosstalk
probabilities of (a) $\epsilon_{nn}=0.078\pm0.001$, $\epsilon_{1stage}=0.5\pm0.3$,
$\epsilon_{rec}=0.06\pm0.02$ (green circles), and (b) $\epsilon_{nn}=0.010\pm0.001$,
$\epsilon_{1stage}=0.1\pm0.1$, $\epsilon_{rec}=0.02\pm0.01$ (blue
squares). Experimental errors are smaller than their respective symbol
size.\label{fig:models comparision}\selectlanguage{british}
}
\end{figure}

In order to demonstrate the use of our model for reconstructing the
original photon statistics from measured data, we present in figure~\ref{fig:models comparision}
thermal photon-number statistics measured using a SiPM detector. The
measurements were conducted on a single polarization mode\foreignlanguage{english}{
from a collinear type-II parametric down-conversion source~\cite{MandelWolf_OpticalCoherence}
with a pulsed pump at a repetition rate of 250\,kHz.} The detector
was operated using two different bias voltage values, which changed
the crosstalk probability between the two measurements. The photon-number
distribution which should\foreignlanguage{english}{ appear as a straight
line in a semi-log plot ($p(n)\sim\left(\frac{\left\langle n\right\rangle }{1+\left\langle n\right\rangle }\right)^{n}$)
experiences a change in slope due to the crosstalk effect. Zero photons
cannot generate crosstalk, and thus the probability of zero photons
is not affected. However, the probability of measuring two or more
photons grows considerably due to crosstalks.} We fit the experimental
data to the photon-number statistics obtained by our computational
model, the one-stage crosstalk model of reference~26 and the recursive
model of reference~28. The crosstalk probability defined in previously
presented models is defined as the overall probability that crosstalk
will be generated, rather than the probability of triggering a specific
neighbour. We associate this crosstalk value with our defined $\epsilon_{nn}$
through the relation $\epsilon=1-\left(1-\epsilon_{nn}\right)^{4}$,
four being the number of nearest neighbours in the square lattice.
When the crosstalk effect is weak, all models fit the data. When the
crosstalk effect is strong, the 1-stage model and the recursive model
show large deviations from the data. An analytical model which provides
a good description for all CT values is currently under work.

\selectlanguage{english}%

\selectlanguage{british}%
In conclusion, we have simulated the detection process in SiPM detectors
and have shown that it is highly affected by the finite number of
detection elements. These effects must be taken into account when
the photon-number statistics of impinging photons is reconstructed.
\foreignlanguage{english}{We have shown that simplistic modeling which
does not account for the geometrical properties of the detection elements
is applicable provided the number of detected photons is below some
threshold which depends on the fraction of triggered elements and
on the crosstalk probability. We have also shown the number of crosstalk
stages which must be taken into account in order to properly model
the crosstalk effect. Finally, we demonstrated the application of
our computational model on experimental data.}

\selectlanguage{english}%

\selectlanguage{british}%

\section*{References}

\bibliographystyle{iopart-num}
\bibliography{references}
\selectlanguage{english}

\end{document}